\documentclass[12pt, reqno, harvard]{amsart}
\usepackage{harvard,graphics,amsmath,amssymb,epsfig,stmaryrd,color}
\usepackage{amsfonts}

\numberwithin{equation}{section}

\begin{document}
\title[Set coverage]{Set coverage and robust policy}
\author{Marc Henry and Alexei Onatski}

\maketitle



When conducting inference on partially identified parameters, \citeasnoun{IM:2004} pointed out that confidence regions may cover the whole identified set with a prescribed probability, to which we will refer as {\em set coverage}, or they may cover each of its point with a prescribed probability, to which we will refer as {\em point coverage}. Since set coverage implies point coverage, confidence regions satisfying point coverage are generally preferred on the grounds that they may be more informative. The object of this note is to describe a decision problem in which, contrary to received wisdom, point coverage is clearly undesirable.

Consider a random vector $s=(X,\varepsilon)$ on $\{1,\ldots,N\}$. Call realizations $s_i$, $i=1,\ldots,N$ of this random vector states of the world, and call their collection $S=\{s_1,\ldots,s_N\}$. Suppose states of the world are partially observable, by which we mean that the realizations $\{x_1,\ldots,x_N\}$ of $X$ can be observed over repeated experiments, but not the realizations $\{\varepsilon_1,\ldots,\varepsilon_N\}$ of $\varepsilon$. Call $P_X$ the probability mass function of random vector $X$. Let $\Theta$ be a set of models for the states, defined by the fact that for each $\theta\in\Theta$, $P_\theta$ denotes a probability mass function for the random vector $(X,\varepsilon)$. The identified set $\Theta_I$ is defined in the following way: \[\Theta_I=\left\{
\theta\in\Theta:\;\sum_{j=1}^NP_\theta(x_i,\varepsilon_j)=P_X(x_i)\mbox{ for }i=1,\ldots,N\right\}.\] More generally, any additional a priori restriction on the joint distribution of $(X,\varepsilon)$ can be incorporated in the definition of the identified set.

Suppose a decision maker may choose among actions in a set $\mathcal A=\{a_1,\ldots,a_K\}$. The actions may be treatments, as in \citeasnoun{Manski:2004} or policy controls as in \citeasnoun{Brainard:67}. Actions in $\mathcal A$ are defined as functions from $S$ to real valued outcomes. Call $U(a,\theta)$ the ex-ante utility of the decision maker, when $P_\theta$ is the true data generating process for $(X,\varepsilon)$. Typically, this will be von Neumann-Morgenstern expected utility $U(a,\theta)=\int a(s)dP_\theta(s)$. We shall consider two robust decision making procedures based on the identified set: (i) maxmin, where the decision maker maximizes the functional evaluation $V(a)=\min_{\Theta_I}U(a,\theta)$ over $\mathcal A$ and (ii) minmax regret, where the decision maker maximizes $V(a)=\min_{\Theta_I}[U(a,\theta)-\max_{a\in\mathcal A}U(a,\theta)]$ over $\mathcal A$. The arguments we make do not depend on which of the two options (i) or (ii) is chosen, so we shall concentrate on a maxmin decision maker.

The decision maker is supposed to have access to two types of confidence regions for $\Theta_I$ based on repeated sampling in the state space. A region covering the identified set called $\Theta_{SC}$ such that $\mathbb P(\Theta_I\in\Theta_{SC})=1-\alpha$ and a region covering each point of the identified set called $\Theta_{PC}$ such that $\min_{\theta\in\Theta_I}\mathbb P(\theta\in\Theta_{PC})=1-\alpha$. Without necessarily subscribing to the learning model of \citeasnoun{ES:2007}, we appeal to \citeasnoun{AHS:2003} and ``in designing a robust decision rule, we assume that our decision maker worries about
alternative models that available data cannot readily dispose of''. Hence, the decision maker considers two decision rules based on the two respective confidence regions. The decision rule based on $\Theta_{SC}$ consists in choosing $\hat a_{SC}$ in $\mathcal A$ to maximize $\min_{\theta\in\Theta_{SC}}U(a,\theta)$ and the decision rule based on $\Theta_{PC}$ consists in choosing $\hat a_{PC}$ in $\mathcal A$ that maximizes $\min_{\theta\in\Theta_{PC}}U(a,\theta)$. The decision rule based on $\Theta_{SC}$ is robust in the sense that \[\mathbb P\left(\min_{\theta\in\Theta_I}U(\hat a_{SC},\theta)\geq\min_{\theta\in\Theta_{SC}}U(\hat a_{SC},\theta)
\right)\geq1-\alpha,\] so that $\min_{\theta\in\Theta_{SC}}U(a,\theta)$ provides a lower bound for the actual utility functional $V(\hat a_{SC})$ with probability at least as large as $1-\alpha$. The decision rule based on $\Theta_{PC}$, however, is not robust as will be shown with the following example that we contrived in the simplest possible way for expositional purposes.

Let $\{1,\ldots,N\}$ be a population of individuals and let $X\in\{F,M\}$ be their gender and $\varepsilon\in\{T,N\}$ be their talent ($T$ for {\em talented} and $N$ for {\em not so talented}). Half the population is male and half the population is talented, but the correlation $\theta$ between talent and gender is unknown. The decision maker is a social planner who can offer an education opportunity to women only (action $a_1$), to men only (action $a_2$) or to everyone (action $a_3$). The net benefit of offering the education opportunity to a talented person is $B$. The net benefit of offering the education opportunity to a not so talented person is $-B$ (wasted resources). The net benefit of failing to offer the education opportunity to a not so talented person is zero. Finally, the net benefit of failing to offer the education opportunity to a talented person is $-B$ (wasted talent). Assume that the parameter set is equal to $\Theta=\{\theta_1,\theta_2,\theta_3,\theta_4\}$, where under $\theta_1$ all talent is male, and under $\theta_2$ all talent is female, under $\theta_3$ everyone is talented and under $\theta_4$ no one is talented. Given the a priori constraints on the joint distribution of gender and talent, the identified set is $\Theta_I=\{\theta_1,\theta_2\}$. Hence, with von Neumann-Morgenstern expected utility, we have:
\begin{eqnarray*}U(a,\theta_1)=
\frac{1}{2}U(a(F,N))+\frac{1}{2}U(a(M,T))=\left\{
\begin{array}{rcl}-B&\mbox{if}&a=a_1\\
B/2&\mbox{if}&a=a_2\\0&\mbox{if}&a=a_3\end{array}\right.\end{eqnarray*}
and \begin{eqnarray*}U(a,\theta_2)=
\frac{1}{2}U(a(F,T))+\frac{1}{2}U(a(M,N))=\left\{
\begin{array}{rcl}B/2&\mbox{if}&a=a_1\\
-B&\mbox{if}&a=a_2\\0&\mbox{if}&a=a_3\end{array}\right.\end{eqnarray*}
Note that $\mathbb P(\theta_1\notin\Theta_{SC}\mbox{ or }\theta_2\notin\Theta_{SC})=\alpha$, whereas for $\Theta_{PC}$ we only require that either $\mathbb P(\theta_1\notin\Theta_{PC})\leq\alpha$ and $\mathbb P(\theta_2\notin\Theta_{PC})=\alpha$, or $\mathbb P(\theta_1\notin\Theta_{PC})=\alpha$ and $\mathbb P(\theta_2\notin\Theta_{PC})\leq\alpha$. If $\Theta_{PC}$ is more informative than $\Theta_{SC}$, then $2\alpha\geq\mathbb P(\theta_1\notin\Theta_{PC}\mbox{ or }\theta_2\notin\Theta_{PC})>\alpha$. Now, if say $\theta_1\notin\Theta_{PC}$ and $\Theta_{PC}=\{\theta_2\}$, then $\min_{\theta\in\Theta_{PC}}U(a,\theta)=B/2$ if $a=a_1$, $-B$ if $a=a_2$ and $0$ if $a=a_3$ and symmetrically if $\theta_2\notin\Theta_{PC}$.
Hence, when $\theta_j\notin\Theta_{PC}$, the action that maximizes $\min_{\theta\in\Theta_{PC}}U(a,\theta)$ is $\hat a_{PC}=a_j$ and $\min_{\theta\in\Theta_I}U(\hat a_{PC},\theta)=\min_{\theta\in\Theta_I}U(a_j,\theta)=-B$ which can be much smaller than $\min_{\Theta_{PC}}U(a_j,\theta)=B/2$. Hence the action taken on the basis of the region with point coverage yields a utility that may be much smaller than it appears with a probability strictly larger than $\alpha$. In contrast $\hat a_{SC}=a_3$ with probability at least $1-\alpha$ so $\min_{\Theta_{SC}}U(\hat a_{SC},\theta)=0$ and $\min_{\Theta_I}U(\hat a_{SC},\theta)=0$ with probability at least $1-\alpha$, so that decision based on the region providing set coverage does not suffer from the same lack of robustness.

\section*{acknowledgements} Financial support
from SSHRC Grant 410-2010-242 is gratefully acknowledged. Correspondence addresses: Alexei Onatski, Faculty of Economics, Cambridge University, Austin Robinson Building, Sidgwick Avenue, Cambridge CB3 9DD, UK, ao319@cam.ac.uk.

\bibliography{Coverage}
\bibliographystyle{mf}

\end{document}